\documentstyle[12pt]{article}
\begin{document}\thispagestyle{empty}\begin{flushright}
OUT--4102--60\\UTAS--PHYS--95--40\\MZ--TH/95--22\\hep-ph/9509296\\
September, 1995
            \end{flushright} \vspace*{2mm} \begin{center} {\large\bf
Unknotting the polarized vacuum of quenched QED$^{*)}$
            } \vglue5mm{\bf
D.~J.~Broadhurst$^{1)}$
            }\vglue 2mm
Physics Department, Open University\\
Milton Keynes MK7 6AA, United Kingdom
            \vglue 4mm{\bf
R.~Delbourgo$^{2)}$
            } \vglue 2mm
Department of Physics, University of Tasmania\\
GPO Box 252C, Hobart, Tasmania 7001, Australia
            \vglue 4mm{\bf
D.~Kreimer$^{3)}$
            } \vglue2mm
Institut f\"{u}r Physik, Johannes Gutenberg-Universit\"{a}t\\
Postfach 3980, D-55099 Mainz, Germany
            \end{center} \vfill
{\bf Abstract}
A knot-theoretic explanation is given for the rationality of the quenched
QED beta function. At the link level, the Ward identity entails
cancellation of subdivergences generated by one term of the skein relation,
which in turn implies cancellation of knots generated by the other term. In
consequence, each bare three-loop diagram has a rational Laurent expansion
in the Landau gauge, as is verified by explicit computation. Comparable
simplification is found to occur in scalar electrodynamics, when computed
in the Duffin-Kemmer-Petiau formalism.
                                                    \vfill
                                                    \footnoterule\noindent
$^*$) Work supported in part by grant CHRX--CT94--0579, from HUCAM,\\
and grant A69231484, from the Australian Research Council.\\
$^1$) D.Broadhurst@open.ac.uk\\
$^2$) Delbourgo@physvax.phys.utas.edu.au\\
$^3$) Kreimer@dipmza.physik.uni-mainz.de
\newpage
\setcounter{page}{1}
\newcommand{\df}[2]{\mbox{$\frac{#1}{#2}$}}
\newcommand{\Eq}[1]{~(\ref{#1})}
\newcommand{\Eqq}[2]{~(\ref{#1},\ref{#2})}
\newcommand{\ep}{\epsilon}
\newcommand{\ze}[1]{\zeta_{#1}}
\newcommand{\sz}{^{[0]}}
\newcommand{\so}{^{[1]}}
\newcommand{\st}{^{[2]}}
\newcommand{\gm}[1]{\gamma_{#1}}
\newcommand{\be}[1]{\beta_{#1}}
\newcommand{\rd}{{\rm d}}
\newcommand{\pslash}{p\llap{/\kern-0.5pt}}
\newcommand{\Section}[1]{\noindent{\large\bf #1}\quad}
\Section{1. Introduction}
The surprising rationality of the three-~\cite{JLR} and
four-loop~\cite{4LQ} quenched (i.e.\ single-electron-loop) terms in the QED
beta function is an outstanding puzzle, which we here elucidate by giving
the Ward identity $Z_1=Z_2$ an interpretation in terms of the skeining
relation that is the basis of the recent association of
knots~\cite{KN1,KN2,KN3,KN4} with transcendental counterterms.

In Section~2, we study the intricate cancellations between transcendentals
in all 6 of the methods (known to us) for calculating $\beta(a)\equiv
\rd a/\rd\ln\mu^2=\sum_{n}\be{n}a^{n+1}$, with a QED coupling
$a\equiv\alpha/4\pi$. Our focus is the rationality of $\be3\so=-2$~\cite{JLR}
and $\be4\so=-46$~\cite{4LQ}. (Subscripts denote the number of loops;
where necessary, superscripts denote the number of electron loops.)
In Section~3, we compound the puzzle by exposing
even more intricate cancellations of $\ze3$ in three-loop scalar
electrodynamics~\cite{3LS}. The argument from knot theory is given in
Section~4, leading to specific predictions, confirmed by detailed
calculations, performed using the techniques of~\cite{DJB,OSM,BKT}.
Conclusions, regarding higher orders~\cite{4LQ}, unquenched (i.e.\
multi-electron-loop) contributions~\cite{DRR,KKO,LNF}, and non-abelian gauge
theories~\cite{TVZ,LM}, are presented in Section~5.

\Section{2. Six calculations in search of an argument}
There are (at least) 6 ways of obtaining the beta function of quenched QED.
For each, we expose the delicate cancellation of transcendentals between
diagrams, which cries out for explanation.

{\em Method 1: Dyson-Schwinger skeleton expansion}~\cite{JLR,JWB}. The
Dyson-Schwinger equations give the photon self-energy, schematically,
as~\cite{BD}
\begin{equation}
\Pi_{\mu\nu}=Z_1\Gamma_\mu G \gamma_\nu
=\Gamma_\mu G (1-KG) \Gamma_\nu\,,
\label{DS}
\end{equation}
where $\Gamma_\mu$ is the dressed vertex and $G$ stands for the pair of
dressed propagators in Fig.~1a. To obtain the second form, illustrated in
Figs.~1b,c, one uses $\Gamma_\nu=Z_1\gamma_\nu+K G\Gamma_\nu$, where $K$ is
the kernel for ${\rm e}^+{\rm e}^-$ scattering and loop integrations
and spin sums are
to be understood in the products. Now we expand each vertex to first order
in the external momentum $q$: $\Gamma_\mu
=\Gamma_\mu^0+q\cdot\rd\Gamma_\mu+O(q^2)$ where $\rd_\alpha\equiv\partial/
\partial q_\alpha$ and the Ward identity gives
$\Gamma_\mu^0\equiv\Gamma_\mu(p,p)= (\partial/\partial p_\mu) S^{-1}(p)$ in
term of the inverse propagator. We use the low-momentum expansion
\begin{equation}
(\Gamma_\mu-\Gamma_\mu^0)G(1-KG)(\Gamma_\nu-\Gamma_\nu^0)
=(q\cdot\rd\Gamma_\mu)G(1-KG)(q\cdot\rd\Gamma_\nu)+O(q^3)\,,
\label{Oq}
\end{equation}
differentiate twice, and make liberal use of
$(1-KG)\rd_\alpha\Gamma_\nu=\rd_\alpha(KG)\Gamma_\nu^ 0+O(q)$, to obtain
\begin{eqnarray}
\rd_\alpha \rd_\beta\Pi_{\mu\nu}
&=&\Gamma_\mu^0\left[
\df{1}{2}\rd_\alpha\rd_\beta G
+(\rd_\alpha G)K(\rd_\beta G)
+2(\rd_\alpha G)(\rd_\beta K)G\right.
+\df{1}{2}G(\rd_\alpha \rd_\beta K)G
\nonumber\\&&\left.{}
+(\rd_\alpha(GK))G(1-KG)^{-1}(\rd_\beta(KG))
+(\alpha\leftrightarrow\beta)\right]\Gamma_\nu^0+O(q)\,,
\label{five}
\end{eqnarray}
which entails only $K$ and $S$. There is a very simple statement of this
result: between the
dressed zero-momentum vertices occur all and only the terms in
$\rd_\alpha\rd_\beta(G(1-KG)^{-1})$ that give no subdivergences. In other
words, the Dyson-Schwinger method kills maximal forests of subdivergences
both on the left and on the right; only the overall divergence survives.
Moreover, after cutting at a line on the left (or right), we may set both
the mass and the external momentum to zero, with no danger of infrared
divergence, and hence obtain the $L$-loop quenched beta function from
finite massless two-point skeleton diagrams, with up to $(L-1)$-loops.

We route half of the external momentum through the electron line and half
through the positron line and find that each of the 5 terms in\Eq{five}
yields a contraction-independent contribution to $\be3\so$. To all orders,
the first contribution to $\beta\so(a)$ is
$\frac43a^2(1-\frac32\gamma^2+\gamma^3)$, where
$\gamma\equiv\gamma_2\sz=\xi a-\frac32a^2+\frac32a^3+O(a^4)$, is the
quenched electron-field anomalous dimension~\cite{OVT} with a photon
propagator $g_{\mu\nu}/k^2+(\xi-1)k_\mu k_\nu/k^4$. The second term gives
$\be2=4$. At 3 loops, all 5 terms contribute, giving a remarkable
cancellation of $\ze3$:
\begin{eqnarray}
\be3\so=-2\xi^2
&+&\left[-\df{2}{3}(7\xi-3)(\xi-3)+2(\xi^2-6\xi-3)\ze3\right]
\,+\,(\xi-1)(\xi-5)\left[\df{16}{3}-4\ze3\right]\nonumber\\
&+&\left[\df{4}{3}(\xi^2+12\xi-23)+2(\xi^2-6\xi+13)\ze3\right]
\,+\,8=-2\,.
\label{b3}
\end{eqnarray}
In the Landau gauge, $\xi=0$, we reproduce Rosner's $\ze3$-cancellation:
$0=-6-20+26$~\cite{JLR}. In the Feynman gauge, $\xi=1$, the third term
vanishes; the second and fourth are still transcendental and exactly cancel
the fifth.

{\em Method 2: Integration by parts of massive bubble diagrams}~\cite{DJB}.
We may study separately each of the 3-loop quenched diagrams of Fig.~2,
using the techniques of~\cite{DJB}, which reduce to algebra the calculation
of 3-loop massive bubble diagrams, in $d\equiv4-2\ep$ dimensions, thereby
also yielding the finite parts of on-shell charge renormalization, used
in~\cite{BKT} to establish the connection between 4-loop on-shell and
minimally subtracted (unquenched) beta functions. The coefficient, $d_n$,
of $a^n/\ep$, in the sum of all $n$-loop, single-electron-loop, bare
diagrams contributing to $1/Z_3$, differs from $\be{n}\so/n$; one must also
take account of the quenched anomalous mass dimension, $\gamma_{\rm m}\sz
=\sum_n\gamma_n a^n=3a+\frac32a^2+\frac{129}{2}a^3+O(a^4)$~\cite{OVT}. To 4
loops, the effects of mass renormalization are also rational, giving
agreement with 3-loop results~\cite{DJB,BKT}:
\begin{equation}
d_1=\be1=\frac43\,;\quad d_2=\df{1}{2}\be2-2\be1\gm1=-6\,;\quad
d_3=\df{1}{3}\be3\so-2\be2\gm1-\be1\gm2+6\be1\gm1^2=\frac{136}{3}\,;
\label{m3}
\end{equation}
and generating a 4-digit prime in the quenched 4-loop coefficient:
\begin{equation}
d_4=\df{1}{4}\be4\so-2\be3\so\gm1-\be2\gm2+8\be2\gm1^2
+8\be1\gm2\gm1-\df{2}{3}\be1\gm3-\df{64}{3}\be1\gm1^3=-\frac{2969}{6}\,,
\label{m4}
\end{equation}
which string-inspired techniques~\cite{SS} may eventually reproduce. In the
simplest case~\cite{DJB} of contracting
$\rd_\alpha\rd_\beta\Pi_{\mu\nu}|_{q=0}$ with $g_{\mu\nu}g_{\alpha\beta}$
in the Feynman gauge, we find cancelling coefficients of $\ze3/\ep$ in the
contributions of Figs.~2f and~2g, corresponding to $\ze3$-cancellation
between the fourth and second terms, respectively, of\Eq{b3} at $\xi=1$.

{\em Method 3: Integration by parts of massless two-point
diagrams}~\cite{CT,A31}.
Next we study the behaviour of the bare diagrams
of Fig.~2 at large $q^2$~\cite{A31},
where we may set $m=0$ and have no need of mass
renormalization, or differentiation. The integration-by-parts method of
Chetyrkin and Tkachov~\cite{CT} now suffices. It has been implemented in
the program {\sc mincer}, whose test suite~\cite{CPC} evaluates all the
diagrams of Fig.~2 in the Feynman gauge. Table~1 of~\cite{CPC} reveals
that, in addition to the crossed-photon diagrams of Fig.~2f and~2g, the
uncrossed-photon diagrams of Figs.~2a and~2e also entail $\ze3/\ep$.
Moreover, the cancellation does not occur in the crossed and uncrossed
sectors separately; the relative weights of $\ze3/\ep$ in the contributions
of the bare diagrams 2a,e,f,g are 3 : $-6$ : $-1$ : 4\,.

{\em Method 4: Infrared rearrangement of massless
diagrams}~\cite{TVZ,GPX,IRA}.
Infrared rearrangement is a technique for reducing the calculation of
$L$-loop counterterms to that of $(L-1)$-loop massless two-point diagrams,
by the subtraction of subdivergences in the MS scheme, followed by
nullification of external momenta and appropriate cutting of massless
bubble diagrams. As shown by Method~1, the Dyson-Schwinger equations make
this unnecessary in QED. However, the technique prospered at the 3-loop
level in QCD~\cite{TVZ,GPX} to such an extent as to encourage the
calculation of the 4-loop beta function of QED, with the rational quenched
result $\be4\so=-46$~\cite{4LQ}. A measure of the seemingly miraculous
cancellation of transcendentals in this method is afforded by the
complicated combination of $\ze3$ and $\ze5$ that occurs in the {\em
non\/}-abelian 4-loop QCD corrections to $R({\rm e}^+{\rm e}^-\to{\rm
hadrons})$~\cite{GKL}.

{\em Method 5: Propagation in a background field}~\cite{JB,CGS}. As
observed in~\cite{JB}, the derivative $(\rd/\rd
a)(\beta\so(a)/a^2)=\sum_{n>1} (n-1)\be{n}\so a^{n-2}$ may be obtained from
the coefficient of $\alpha F_{\mu\nu}^2$ in the single-electron-loop
contributions to the large-$q^2$ photon propagator in a background field
$F_{\mu\nu}$, thereby providing a further alternative to infrared
rearrangement. Two-loop massless background-field calculations have been
performed in QCD~\cite{CGS,ST}, with results recently confirmed by a full
analysis of the massive case~\cite{GG2}. {}From the two-loop corrections,
$1+(2C_A-C_F)\alpha_s/4\pi$, to the coefficient of ${<}\alpha_s
G_{\mu\nu}^2{>}$ in the correlator of the light-quark vector current of
QCD, one immediately obtains $2\be3\so/\be2=-1$, by setting $C_A=0$,
$C_F=1$, which confirms the correctness of this method at 3 loops. A
measure of the complexity of the $\ze3$ cancellations is afforded by
studying the diagram-by-diagram analysis of the appendix of~\cite{ST}.

{\em Method 6: Crewther connection to deep-inelastic
processes}~\cite{CBK,DIS}. Finally, a useful check of $\be4\so=-46$ was
obtained in~\cite{CBK} by taking the reciprocal of the 3-loop~\cite{DIS}
radiative corrections to the Gross--Llewellyn-Smith sum rule, in the
quenched abelian case:
\begin{equation}
\beta\so(a)=\frac{\df{4}{3}a^2}{1-3a+\df{21}{2}a^2-\df32{a^3}
+O(a^4)}=\df{4}{3}a^2+4a^3-2a^4-46a^5+O(a^6)\,,
\label{cbk}
\end{equation}
in precise agreement with~\cite{4LQ}. This is an example of a Crewther
connection~\cite{RJC} which is unmodified by renormalization in quenched
QED. A measure of the complexity of the cancellations of $\ze3$ and $\ze5$
is afforded by studying the {\em non\/}-abelian terms in 3-loop
deep-inelastic radiative corrections~\cite{DIS}, which are replete with
both transcendentals.

\Section{3. Cancellation of $\ze3$ in scalar electrodynamics}
Corresponding calculations in scalar electrodynamics (SED) are most
conveniently performed using the Duffin-Kemmer-Petiau~\cite{DKP} (DKP)
spinorial formalism for the charged scalar field, in which the Feynman
rules are identical to those of fermionic QED, with a bare vertex
$e_0\gamma_\mu$ and a bare propagator $S_0(p)=1/(\pslash-m_0)$. The only
difference resides in the $\gamma$-matrices: in the DKP formalism $\pslash$
is not invertible; instead one uses the fact that $\pslash^3=p^2\pslash$ to
obtain $S_0(p)=(\pslash(\pslash+m_0)/(p^2-m_0^2)-1)/m_0$. The trace of the
unit matrix is $d+1$, the trace of an odd number of $\gamma$-matrices
vanishes, and the trace of an even number, $\gamma_{\mu_1}$ to
$\gamma_{\mu_{2n}}$, is the sum of two terms~\cite{JMP}:
$g_{\mu_1\mu_2}\ldots g_{\mu_{2n-1}\mu_{2n}}$ and the cyclic permutation
$g_{\mu_2\mu_3}\ldots g_{\mu_{2n}\mu_1}$. The effect is to make all traces
regular as $m_0\to0$, while automatically generating the many seagull terms
of conventional scalar methods~\cite{BD}. This is a great simplification,
eliminating the need to include 21 seagull diagrams at 3 loops, which would
lead to many more terms in $\rd_\alpha\rd_\beta\Pi_{\mu\nu}|_{q=0}$.

We compute the diagrams of Fig.~2 using the DKP formalism. At $q=0$ (Method
2) we use the {\sc reduce}~\cite{RED} program {\sc recursor}~\cite{DJB},
for 3-loop massive bubble diagrams; at $m=0$, we use the {\sc reduce}
program {\sc slicer}~\cite{BKT}, devised to check the results of~\cite{A31}
for the large-$q^2$ photon propagator in the $\overline{\rm MS}$ scheme. As
in QED, we perform both calculations in an arbitrary gauge and contract
$\rd_\alpha\rd_\beta\Pi_{\mu\nu}$ with
$g_{\mu\nu}g_{\alpha\beta}+\lambda
(g_{\mu\nu}g_{\alpha\beta}+g_{\mu\alpha}g_{\nu\beta}
+g_{\nu\alpha}g_{\mu\beta})$, where $\lambda$ is an arbitrary parameter,
which affects the contributions of individual diagrams, but not the total
result for the transverse self energy $\Pi_{\mu\nu}(q)=(q_\mu q_\nu
-q^2g_{\mu\nu})\Pi(q^2)$. At $m=0$, the corresponding freedom is to
contract with the tensor $g_{\mu\nu}q^2+\lambda(d+2) q_\mu q_\nu$. On-shell
SED mass renormalization is performed as in~\cite{DJB,OSM}: there are only
two quenched self-energy diagrams at two loops; each is projected on-shell
by taking the trace with $\pslash( \pslash+m)$ at $p^2=m^2$, with a pole
mass $m$. Unlike the QED case, the relation between bare and pole masses is
infrared-singular in SED, though that causes no problem for the
dimensionally regularized calculation of $Z_3=1/(1+\Pi_0(0))$, where
infrared singularities in the bare diagrams for $\Pi_0(0)$ are cancelled by
those in $Z_{\rm m}\equiv m_0/m$. The on-shell methods of~\cite{DJB} then
yield the quenched contributions
\begin{equation}\frac{1}{Z_3\so}-1=\left\{
\begin{array}{l}
\frac{4}{3\ep}a_{\rm m}
+\frac{4(1+7\ep-4\ep^3)}{\ep(2-\ep)(1-4\ep^2)}a_{\rm m}^2
+\left[-\frac{2}{3\ep}
+16\ze2(5-8\ln2)+\frac13\ze3+\frac{77}{9}\right]a_{\rm m}^3,\\[3pt]
\frac{1}{3\ep}a_{\rm m}+\frac{4}{\ep(2-\ep)(1-4\ep^2)}a_{\rm m}^2
+\left[+\frac{29}{6\ep}
+8\ze2(3-4\ln2)+\frac{35}{12}\ze3+\frac{136}{9}\right]a_{\rm m}^3,
\end{array}\right.\label{Z3}
\end{equation}
for QED~\cite{DJB,BKT} and SED, respectively, where $a_{\rm
m}=\Gamma(1+\ep)e_0^2/(4\pi)^{d/2} m^{2\ep}$ is a dimensionless coupling,
with a pole mass $m$, and terms of order $a_{\rm m}^3\ep$ and $a_{\rm m}^4$
are neglected. (There is no need to renormalize the bare charge, $e_0$, when
dealing with the quenched contributions.)
{}From the singular terms in\Eq{Z3} we read off the 3-loop beta function of
quenched SED: $\tilde{\beta}\so(a)=\frac13a^2+4a^3+\frac{29}{2}a^4+O(a^5)$.
We have calculated the double-bubble term $\tilde{\beta}_3\st$ in the
on-shell and MS schemes, obtaining agreement with~\cite{3LS} in the latter,
as shown in Table~1.

For the 3-loop quenched $\overline{\rm MS}$ contributions to $\Pi(q^2)$,
at large $q^2$, we obtain
\begin{equation}\overline{\Pi}_3\so(q^2)=
\left\{\begin{array}{l}
\left[-2\ln(\mu^2/q^2)-\frac{286}{9}-\frac{296}{3}\ze3+160\ze5
\right]\bar{a}^3,\\[3pt]
\left[+\frac{29}{2}\ln(\mu^2/q^2)+\frac{502}{9}-\frac{160}{3}\ze3+40\ze5
\right]\bar{a}^3,
\end{array}\right.\label{Pi3}
\end{equation}
with $\bar{a}=\Gamma(1+\ep)\Gamma^2(1-\ep)e_0^2/
\Gamma(1-2\ep)(4\pi)^{d/2}\mu^{2\ep}$, which suppresses $\ze2$ in Laurent
expansions. The QED result confirms~\cite{A31}. The SED
cancellations are even subtler:
with $\xi=\lambda=0$ the relative weights of $\ze3/\ep$ from
bare diagrams 2a,e,f,g,h are 18 : $-36$ : $-5$ : 22 : 1\,.

\Section{4. The argument from knot theory}
{}From the point of view of knot theory, as proposed
in~\cite{KN1,KN2,KN3,KN4}, the presence of transcendentals in counterterms
can be traced to the knots that are obtained by skeining the link diagrams
that encode the intertwining of loop momenta in Feynman diagrams. The
absence of transcendentals in the quenched beta function does not,
therefore, correspond to the absence of knots in the Feynman graphs, since
the crossed-photon graphs of Figs.~2f,g,h all realize the link diagram
whose skeining contains the trefoil knot~\cite{KN2}. Accordingly we expect
to find $\ze3/\ep$ in their divergent parts. To explain the cancellation
of transcendentals, we must study the interplay between knot-theoretic
arguments and the gauge structure of QED.

It was found in~\cite{KN1,KN2} that ladder topologies are free of
transcendentals when the appropriate counterterms are added: after
minimal subtraction of subdivergences, ladder graphs, such
as in Figs.~2a,e, give rational terms in the Laurent expansion in powers of
$1/\ep$. In~\cite{KN3,KN4}, on the other hand, transcendentals
corresponding to positive knots, with up to 11 crossings, were successfully
associated with subdivergence-free graphs, up to 7 loops. Thus the skein
relation played two distinct roles in previous applications: in~\cite{KN1}
the so-called $A$ part of the skein operation determined the
subdivergences, while the $B$ part gave no non-trivial knots in ladder
topologies; in~\cite{KN3} there were no subdivergences associated with the
$A$ operation, while the knots from the $B$ operation faithfully revealed
the transcendentals resulting from nested subintegrations. In Figs.~2f,g,h
we are now confronted with Feynman diagrams whose link diagrams generate
the trefoil knot (via $B$) and also have subdivergences (corresponding to
$A$).

We thus propose to associate the cancellation of transcendentals with the
cancellation of subdivergences in the quenched beta function of QED, which
is an immediate consequence of the Ward identity, $Z_1=Z_2$.

To see the key role of the Ward identity, consider Fig.~2g. There is an
internal vertex correction, which is rendered local by adding the
appropriate counterterm graph. Due to the Ward identity, this counterterm
graph is the same as that which compensates for the self-energy correction
in Fig.~2e. In~\cite{KN1,KN2} it was shown that the latter counterterm
could be interpreted as the $A$ part of the skein operation on
the link diagram $L(2{\rm e})$ of Fig.~3, associated with the Feynman
graph~2e. We assume that this is a generic feature of the relationship
between skeining and renormalization and associate the corresponding
counterterm for Feynman graph~2g with the term obtained from applying
$A$ twice to the link diagram $L(2{\rm g})$, which
requires two skeinings to generate the same
counterterm, along with the trefoil knot from the $B$ term.
The Ward identity thus becomes a relation between
crossed and uncrossed diagrams, after skeining:
\begin{equation}
A(A(L(2{\rm g})))=A(L(2{\rm e}))\quad\Rightarrow\quad
A(L(2{\rm g}))=L(2{\rm e})\,.
\label{Ward}
\end{equation}
The trefoil knot results, in this language, from
$B(B(L(2{\rm g})))$, which generates, in general, a $\ze3/\ep$ term
from Fig.~2g, even after the subtraction of subdivergences.
We now use the Ward identity\Eq{Ward} at the link
level to obtain
\begin{equation}
B(B(L(2{\rm g}))) = B(B(A^{-1}(L(2{\rm e}))))\,,
\label{arg}
\end{equation}
from which we see that it relates the transcendental
counterterm from a torus-knot~\cite{KN2} topology, in Fig.~2g, to a
knot-free~\cite{KN1} ladder topology, in Fig.~2e.

We conclude that $\ze3/\ep$ should be absent from the bare diagram of
Fig.~2g when it is calculated in the gauge where the bare
diagram of Fig.~2e is free of $\ze3/\ep$, i.e.\ in the Landau
gauge, where the latter is free of subdivergences. In Fig.~3 we summarize
the argument. By a similar argument, we also conclude that the other graphs
with the trefoil topology, namely Figs.~2f,h, should be free of
$\ze3/\ep$ in the Landau gauge.

Note that vanishing of the bare ladder diagram of Fig.~2e, in the Landau
gauge, does not imply the vanishing of $B(B(A^{-1}(L(2{\rm e}))))$; it
merely implies its rationality. The action of the $A$ and $B$
operators is non-trivial on the diagram of
Fig.~2e; in the language of~\cite{KN1,KN2}, skeining involves a
change of writhe number in the subdivergence, and only terms with
writhe number zero vanish in the Landau gauge, while the double
application of the $B$ operator in $B(B(A^{-1}(L(2{\rm e}))))$
generates a non-vanishing writhe for the subdivergence.

To test these ideas, we evaluate the separate contributions of the QED
diagrams of Fig.~2 to\Eqq{Z3}{Pi3} in the Landau gauge and find, indeed,
that {\em each\/} bare diagram has a rational $1/\ep$ term in its Laurent
expansion, at $\xi=0$, for any value of the contraction parameter,
$\lambda$. In any other gauge, the bare massless diagrams of Figs~2a,e,f,g
give non-zero, mutually cancelling, coefficients of $\ze3/\ep$.
The $m=0$ contributions to the
Laurent expansion of $\Pi_3\so$ are given in Table~2, in the case
$\xi=\lambda=0$. The same rational behaviour is observed in the Landau
gauge at $q=0$ (again for any value of $\lambda$) with contributions also
recorded in Table~2 (with $\lambda=0$), along with that from mass
renormalization of lower-loop diagrams.

To determine the corresponding gauge for SED, we study both of the two-loop
DKP diagrams and find that each gives $1/\ep^2$ terms, proportional to
$2\xi-3$. Thus the gauge $\xi=3/2$ is the closest DKP analogy to the Landau
gauge of QED. In any gauge, however, diagram~2h gives a $\ze3/\ep$ term,
proportional to $\lambda-2$. (The absence of such a term in QED is
attributable to the fact that the finite part of the two-loop fermion
propagator does not involve $\ze3$ at large momentum.) Accordingly we
evaluate all $m=0$ and $q=0$ SED diagrams with $\xi=3/2$ and $\lambda=2$
and find that every one is indeed rational, with a Laurent expansion
recorded in Table~3. Again, we regard this as a significant success of
skeining arguments, since fixing 2 parameters removes $\ze3/\ep$ from each
of the 5 diagrams~2a,e,f,g,h.

\Section{5. Conclusions}
Complete cancellation of transcendentals from the beta function, at every
order, is to be expected only in quenched QED and quenched SED, where
subdivergences cancel between bare diagrams.

Contributions with more than one charged loop require scheme-dependent
coupling-constant renormalization. Knot theory guarantees that
double-bubble contributions are rational at the three-loop level, as found
in~\cite{3LS,DRR} and shown in Table~1. But at four
loops~\cite{4LQ,BKT,KKO},
and beyond~\cite{LNF}, such multi-electron-loop contributions entail
non-trivial knots, associated with zeta functions~\cite{KN2} and more
exotic~\cite{KN3,KN4} transcendentals, whose cancellation is {\em not\/}
underwritten by the Ward identity.

Likewise, we do not expect the beta functions of {\em non\/}-abelian gauge
theories to be rational beyond the presently computed~\cite{TVZ,LM}
three-loop level, since coupling-constant renormalization is required to
remove subdivergences generated by gluon- and ghost-loops in the gluon
self-energy, which cannot be quenched without violating the Slavnov-Taylor
identities. Indeed, it is still somewhat obscure how the rationality of
$\be3$ comes about in QCD: as shown in~\cite{LM}, the renormalization
constants $Z_{1,2,3}$, for the quark-gluon vertex, quark field, and gluon
field, all involve $\ze3/\ep$ at three-loops; only in
$Z_{\alpha}=Z_1^2Z_2^{-2}Z_3^{-1}$ does the cancellation occur.
It may be possible
to understand this in a three-loop background-field calculation, but we do
not expect the intrinsically scheme-dependent non-abelian beta function to
remain rational at higher orders in the MS scheme.

In fact, there are only two further quantities that we expect to be
rational beyond three loops: the quenched anomalous field and mass
dimensions, $\gamma_2\sz$ and $\gamma_{\rm m}\sz$, of QED, whose
rationality at the three-loop level is established in~\cite{LM,OVT}. These
are scheme-independent, since subdivergences cancel in the electron
propagator, by virtue of the Ward identity, which here relates proper
self-energy diagrams to one-particle-reducible diagrams.

One is tempted to seek a calculational method, based on the skein relation,
to compute the rational anomalous dimensions of quenched QED, in terms of
diagrams, with non-zero writhe numbers, that have been `ladderized'
by Ward identities, such as\Eq{arg}, which
may eventually yield a rational calculus like that in~\cite{KN2}.
The complexity of the rational contributions of Tables~2 and~3 indicates
that such a calculus would be rather non-trivial.

In conclusion: all-order rationality of counterterms is specific to
quenched abelian theories, where the cancellation of knots, from the $B$
term~\cite{KN3,KN4} in the skein relation, matches the cancellation of
subdivergences, from the $A$ term~\cite{KN1,KN2}.  This is exemplified by
the rationality of all quenched three-loop bare QED diagrams,
in the Landau gauge.
\newpage
\Section{Acknowledgements}
DJB and DK thank the organizers and participants of the Pisa and Aspen
multi-loop workshops for their interest in and comments on this work.
Detailed discussions with Kostja Chetyrkin, John Gracey, Sergei Larin,
Christian Schubert and Volodya Smirnov were most helpful. DJB thanks Andrei
Kataev and Eduardo de Rafael, for long-standing dialogues on the rationality
of quenched QED, and the organizers of the UK HEP Institute in Swansea,
where the computations were completed.

\raggedright

\newpage

Table~1: Three-loop on-shell and MS coefficients of QED and SED
beta functions.
\[\begin{array}{|c|c|c|}
\hline\be3\so+\be3\st&{\rm QED}&{\rm SED}\\[2pt]\hline
\mbox{on-shell}&
-2-\frac{224}{9}=-\frac{242}{9}~\cite{DRR}&
\frac{29}{2}-\frac{55}{18}=\frac{103}{9}~\phantom{[99]}\\[2pt]\hline
\mbox{MS}&
-2-\frac{44}{9}=-\frac{62}{9}\,{}~\cite{TVZ}&
\frac{29}{2}-\frac{49}{18}=\frac{106}{9}~\cite{3LS}\\[2pt]\hline
\end{array}\]
\vfill
Table~2: QED contributions to $\Pi_3\so$, at $m=0$ and $q=0$,
with $\xi=\lambda=0$.
\[\begin{array}{|c|rr|rr|}
\hline\mbox{method}& &m=0& &q=0\\\hline
\mbox{Laurent term}&
\bar{a}^3/\ep^2&
\bar{a}^3/\ep&
a_{\rm m}^3/\ep^2&
a_{\rm m}^3/\ep\\\hline
\mbox{diagram}~2\mbox{a}&
0&
8/3&
0&
8/3\\
\mbox{diagram}~2\mbox{b}&
0&
0&
-12/5&
152/5\\
\mbox{diagram}~2\mbox{c}&
0&
0&
72/5&
8/5\\
\mbox{diagram}~2\mbox{d}&
0&
0&
2/3&
31/3\\
\mbox{diagram}~2\mbox{e}&
0&
0&
0&
-24\\\hline
\mbox{diagram}~2\mbox{f}&
-2/3&
7/3&
-2/3&
17/3\\
\mbox{diagram}~2\mbox{g}&
0&
-28/3&
0&
-28/3\\
\mbox{diagram}~2\mbox{h}&
2/3&
11/3&
-12&
28\\\hline
\mbox{uncrossed}&
0&
8/3&
38/3&
21\\
\mbox{crossed}&
0&
-10/3&
-38/3&
73/3\\
\mbox{mass renorm}&
0&
0&
0&
-46\\\hline
\mbox{total}&
0&
-2/3&
0&
-2/3\\\hline
\end{array}\]
\vfill
Table~3: SED contributions to $\Pi_3\so$, at $m=0$ and $q=0$,
with $\xi=3/2$ and $\lambda=2$.
\[\begin{array}{|c|rrr|rrr|}
\hline\mbox{method}& &&m=0& &&q=0\\\hline
\mbox{Laurent term}&
\bar{a}^3/\ep^3&
\bar{a}^3/\ep^2&
\bar{a}^3/\ep&
a_{\rm m}^3/\ep^3&
a_{\rm m}^3/\ep^2&
a_{\rm m}^3/\ep\\\hline
\mbox{diagram}~2\mbox{a}&
1/16&
55/96&
4151/576&
1/16&
7/96&
2501/576\\
\mbox{diagram}~2\mbox{b}&
-13/8&
-409/24&
-16243/144&
-13/8&
-233/120&
-8/45\\
\mbox{diagram}~2\mbox{c}&
3/2&
31/2&
3703/36&
3/2&
43/20&
203/45\\
\mbox{diagram}~2\mbox{d}&
1/16&
103/96&
4387/576&
1/16&
-65/96&
-131/576\\
\mbox{diagram}~2\mbox{e}&
0&
-5/48&
41/32&
0&
-5/48&
-1141/96\\\hline
\mbox{diagram}~2\mbox{f}&
-1/16&
-9/32&
911/192&
-1/16&
7/32&
1013/192\\
\mbox{diagram}~2\mbox{g}&
0&
3/16&
-917/288&
0&
3/16&
-1349/288\\
\mbox{diagram}~2\mbox{h}&
1/16&
3/32&
-1667/576&
1/16&
-29/32&
-809/576\\\hline
\mbox{uncrossed}&
0&
0&
37/6&
0&
-1/2&
-55/16\\
\mbox{crossed}&
0&
0&
-4/3&
0&
-1/2&
-13/16\\
\mbox{mass renorm}&
0&
0&
0&
0&
1&
109/12\\\hline
\mbox{total}&
0&
0&
29/6&
0&
0&
29/6\\\hline
\end{array}\]
\newpage
\setlength{\unitlength}{0.014cm}
\newbox\shell
\newcommand{\lbl}[3]{\put(#1,#2){\makebox(0,0)[b]{$#3$}}}
\newcommand{\dia}[1]{\setbox\shell=\hbox{
\begin{picture}(200,300)(-100,-150)#1\end{picture}}\dimen0=\ht
\shell\multiply\dimen0by7\divide\dimen0by16\raise-\dimen0\box\shell}
\newcommand{\blob}{\circle*{20}}
\noindent Fig.~1:
Illustration of terms in the Dyson-Schwinger equation~(1).\\
\hspace*{0.5cm}
\dia{
\put(0,0){\oval(120,100)[r]}
\put(0,0){\oval(120,100)[l]}
\put(-60,0){\line(-1,0){20}}
\put(+60,0){\line(+1,0){20}}
\put(0,+50){\blob}
\put(0,-50){\blob}
\put(-60,0){\blob}
\lbl{0}{+100}{Z_1\Gamma_\mu G\gamma_\nu}
\lbl{150}{100}{=}
\lbl{0}{-110}{{\bf a}}
\lbl{40}{-10}{Z_1}}
\hspace{1cm}
\dia{
\put(0,0){\oval(120,100)[r]}
\put(0,0){\oval(120,100)[l]}
\put(-60,0){\line(-1,0){20}}
\put(+60,0){\line(+1,0){20}}
\put(0,+50){\blob}
\put(0,-50){\blob}
\put(+60,0){\blob}
\put(-60,0){\blob}
\lbl{0}{+100}{\Gamma_\mu G\Gamma_\nu}
\lbl{150}{+100}{-}
\lbl{0}{-110}{{\bf b}}}
\hspace{1.5cm}
\dia{
\lbl{0}{-10}{K}
\put(0,0){\oval(60,130)[t]}
\put(0,0){\oval(60,130)[b]}
\put(-30,0){\oval(120,100)[l]}
\put(+30,0){\oval(120,100)[r]}
\put(-90,0){\line(-1,0){20}}
\put(+90,0){\line(+1,0){20}}
\put(-50,+50){\blob}
\put(-50,-50){\blob}
\put(+50,+50){\blob}
\put(+50,-50){\blob}
\put(+90,0){\blob}
\put(-90,0){\blob}
\lbl{0}{+100}{\Gamma_\mu G K G\Gamma_\nu}
\lbl{0}{-110}{{\bf c}}}\\[10pt]
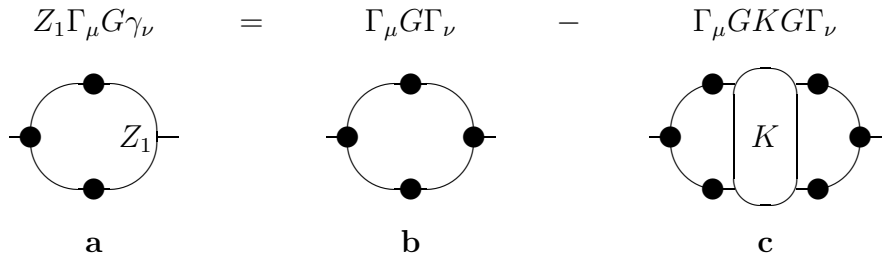
Fig~2: Diagrams contributing to $\beta_3^{[1]}$, with photon lines
drawn inside the electron loop.\\[-10pt]
\hspace*{0.5cm}
\dia{
\put(0,0){\circle{100}}
\put(-50,0){\line(-1,0){20}}
\put(+50,0){\line(+1,0){20}}
\put(-30,-40){\line(0,+1){80}}
\put(+30,-40){\line(0,+1){80}}
\lbl{0}{-110}{{\bf a}}}
\hspace{0.5cm}
\dia{
\put(0,0){\circle{100}}
\put(-50,0){\line(-1,0){20}}
\put(+50,0){\line(+1,0){20}}
\put(-47,47){\oval(60,60)[br]}
\put(+47,47){\oval(60,60)[bl]}
\lbl{0}{-110}{{\bf b}}}
\hspace{0.5cm}
\dia{
\put(0,0){\circle{100}}
\put(-50,0){\line(-1,0){20}}
\put(+50,0){\line(+1,0){20}}
\put(0,+40){\oval(60,60)[b]}
\put(0,-40){\oval(60,60)[t]}
\lbl{0}{-110}{{\bf c}}}
\hspace{0.5cm}
\dia{
\put(0,0){\circle{100}}
\put(-50,0){\line(-1,0){20}}
\put(+50,0){\line(+1,0){20}}
\put(0,40){\oval(60,60)[b]}
\put(0,47){\oval(30,30)[b]}
\lbl{0}{-110}{{\bf d}}}\\[-30pt]
\hspace*{0.5cm}
\dia{
\put(0,0){\circle{100}}
\put(-50,0){\line(-1,0){20}}
\put(+50,0){\line(+1,0){20}}
\put(-47,47){\oval(60,60)[br]}
\put(0,50){\line(0,-1){100}}
\lbl{0}{-110}{{\bf e}}}
\hspace{0.5cm}
\dia{
\put(0,0){\circle{100}}
\put(-50,0){\line(-1,0){20}}
\put(+50,0){\line(+1,0){20}}
\put(-30,-40){\line(+3,+4){60}}
\put(-30,+40){\line(+3,-4){27}}
\put(+30,-40){\line(-3,+4){27}}
\lbl{0}{-110}{{\bf f}}}
\hspace{0.5cm}
\dia{
\put(0,0){\circle{100}}
\put(-50,0){\line(-1,0){20}}
\put(+50,0){\line(+1,0){20}}
\put(0,+50){\line(0,-1){37}}
\put(0,-50){\line(0,+1){56}}
\put(0,+40){\oval(60,60)[b]}
\lbl{0}{-110}{{\bf g}}}
\hspace{0.5cm}
\dia{
\put(0,0){\circle{100}}
\put(-50,0){\line(-1,0){20}}
\put(+50,0){\line(+1,0){20}}
\put(-46,+20){\line(+4,+1){76}}
\put(+46,+20){\line(-4,+1){42}}
\put(-30,+39){\line(+4,-1){26}}
\lbl{0}{-110}{{\bf h}}}\\[10pt]
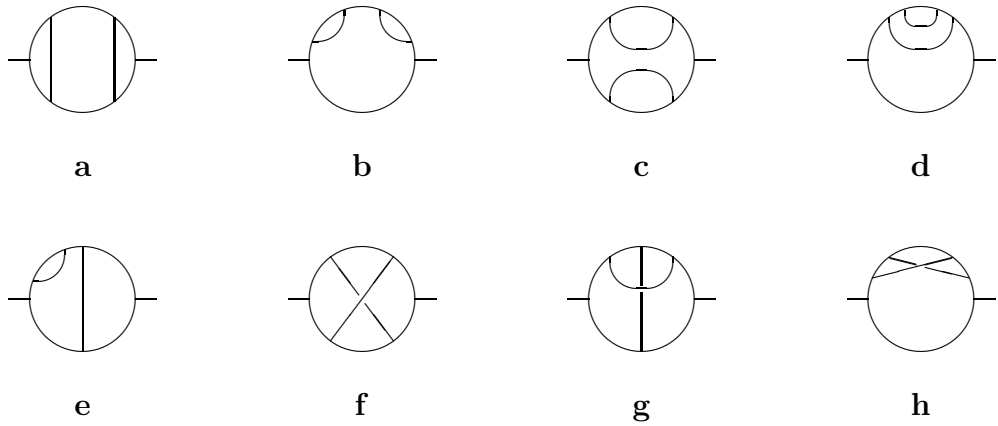
Fig.~3:  The $A$ part of the skein operation on the link diagrams for
Figs.~2e,g.\\[-10pt]
\hspace*{0.5cm}
\dia{
\put(0,0){\circle{100}}
\put(-50,0){\line(-1,0){20}}
\put(+50,0){\line(+1,0){20}}
\put(-47,47){\oval(60,60)[br]}
\put(0,50){\line(0,-1){100}}
\lbl{0}{-110}{{2{\rm e}}}}
\hspace{1mm}
$\sim$
\hspace{5mm}
\dia{
\put(-30,0){\circle{100}}
\put(30,0){\circle{100}}
\put(-50,40){\circle{60}}
\lbl{0}{-110}{{L(2{\rm e})}}}
\hspace{5mm}
$\rightarrow$
\hspace{8mm}
\dia{
\put(-30,0){\circle{100}}
\put(30,0){\circle{100}}
\put(-110,40){\circle{60}}
\lbl{0}{-110}{{A(L(2{\rm e}))}}}\\[-10pt]
\hspace*{0.5cm}
\dia{
\put(0,0){\circle{100}}
\put(-50,0){\line(-1,0){20}}
\put(+50,0){\line(+1,0){20}}
\put(0,+50){\line(0,-1){37}}
\put(0,-50){\line(0,+1){56}}
\put(0,+40){\oval(60,60)[b]}
\lbl{0}{-110}{{2{\rm g}}}}
\hspace{1mm}
$\sim$
\hspace{5mm}
\dia{
\put(-30,0){\circle{100}}
\put(30,0){\circle{100}}
\put(0,37){\circle{60}}
\lbl{0}{-110}{{L(2{\rm g})}}}
\hspace{5mm}
$\rightarrow$
\hspace{8mm}
\dia{
\put(-30,0){\circle{100}}
\put(30,0){\circle{100}}
\put(-50,40){\circle{60}}
\lbl{0}{-110}{{A(L(2{\rm g}))=L(2{\rm e})}}}

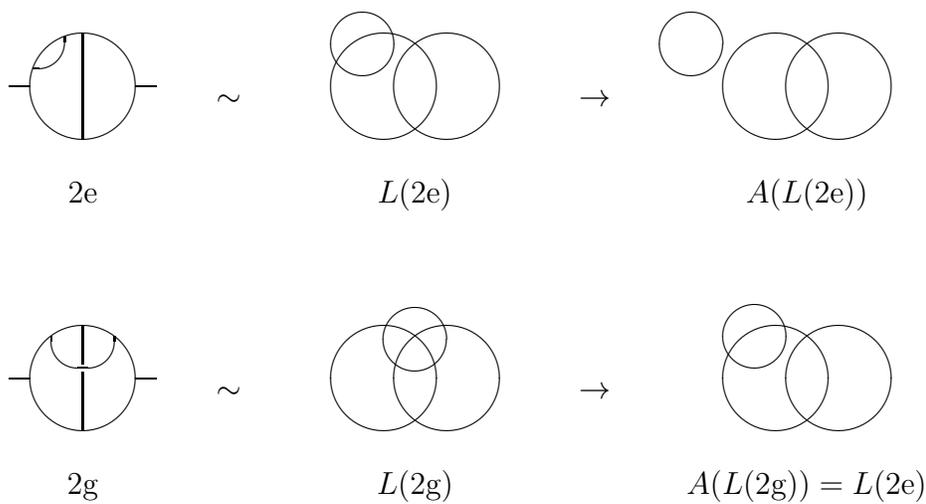
\end{document}